\documentclass[aps,prb,twocolumn,groupedaddress,floatfix,showpacs]{revtex4}

\usepackage{graphics}
\usepackage{epsfig}
\usepackage[latin1]{inputenc}
\usepackage{subfigure}
\usepackage{dcolumn}
\usepackage{bm}
\usepackage{overpic} 

\begin{document}

\title{Size distributions of chemically synthesized Ag nanocrystals}
\author{Annett Th\o gersen}
\author{Jack Bonsak}
\author{Carl Huseby Fosli}
\affiliation{Department of Solar Energy, Institute for Energy Technology, Instituttveien 18, 2007 Kjeller, Norway}

\author{Georg Muntingh}
\affiliation{Centre of Mathematics for Applications, University of Oslo, Norway}

\date{\today}

\begin{abstract}

Silver nanocrystals made by a chemical reduction of silver salts (AgNO$_3$) by sodium borohydride (NaBH$_4$) were studied using Transmission Electron Microscopy (TEM) and light scattering simulations. For various AgNO$_3$/NaBH$_4$ molar ratios, the size distributions of the nanocrystals were found to be approximately log-normal. In addition, a linear relation was found between the mean nanocrystal size and the molar ratio. In order to relate the size distribution of Ag nanocrystals of the various molar ratios to the scattering properties of Ag nanocrystals in solar cell devices, light scattering simulations of Ag nanocrystals in Si, SiO$_2$, SiN, and Al$_2$O$_3$ matrices were carried out using Mie Plot. These light scattering spectra for the individual nanocrystal sizes were combined into light scattering spectra for the fitted size distributions. The evolution of these scattering spectra with respect to an increasing mean nanocrystal size was then studied. From these findings, it is possible to find the molar ratio for which the corresponding nanocrystal size distribution has maximum scattering at a particular wavelength in the desired matrix.

\end{abstract}

\maketitle

\section{Introduction}

Thin Si wafers are attractive for solar cell applications, because they can reduce the material consumption and thereby decrease the cost of solar electricity. However, when a Si wafer decreases in thickness, much of the light passes through the wafer and will not be absorbed. In order to capture the transmitted light, light trapping structures at the front or back side can be incorporated in the solar cell structure.

One way to capture more light is to deposit metallic nanoparticles on the solar cell. Metallic nanoparticles can scatter light into the solar cell by the localized surface plasmon resonance, which is a collective oscillation of the conduction electrons \cite{intro:1}. This strong absorption band is not present in the spectrum of the corresponding bulk metal. The valence electron oscillation results in a local electromagnetic field at the nanocrystal surface and in wavelength selective photon absorption and scattering \cite{intro:1}. By dispersing metal nanocrystals on the surface of an optically thin solar cell, the localized surface plasmon resonance will scatter the light further into the solar cell, thereby increasing the absorption and efficiency. This increase in absorption in wafer-based silicon solar cells has been shown by both Pillai \cite{intro:31} and Beck et al. \cite{intro:beck}. In particular, Pillai et al. \cite{intro:31} showed a 33\% increase of the total current of a device with Ag nanocrystals.  Whereas, Yoon et al. \cite{Woo:Ag} have shown an enhanced optical absorption and an improved J$_{sc}$ for polymer-fulleren bulk hetrojunction PV device. 

The nanocrystals in this work have been made by a chemical synthesis of metallic nanoparticles, using chemical reduction of silver salts by sodium borohydride. This method has shown to be a simple, economical, and popular method that can be applied in large scales as required for industrial applications\cite{intro:2,jack:emrs}. The surface plasmon resonance depends on the shape of the nanocrystal surface, size, spatial arrangement, and configuration of the nanocrystals \cite{intro:4}. It is therefore very important to investigate these features in detail. In our previous work we studied the nanocrystals shape depending on defects in the nanocrystals \cite{annett:mrs}. In this article, the size distribution of nanocrystals made with different molar ratios of AgNO$_3$ to NaBH$_4$ were examined in detail and related to the scattering of light when incorporated into the solar cell device.

\section{Experimental}

The Ag nanocrystals were made by a wet chemical reduction synthesis mixing a silver nitrate solution (AgNO$_3$, 99.9\% purity, Qualigens Fine Chemicals) with a highly reducing solution sodium borohydride (NaBH$_4$, 95\% purity, Merck) without further purification. All equipment was thoroughly cleaned by soaking in ethanol and washed with distilled water. A 0.6 mM AgNO$_3$ precursor solution was prepared using deionized H$_2$O, together with a 1.2 mM aqueous NaBH$_4$ solution. In order to prevent agglomeration, the solutions were kept ice cold and the mixture was stirred continuously. Ag nanocrystals were formed by adding the aqueous AgNO$_3$ drop by drop to the highly reducing solution of NaBH$_4$. Variations in particle size and distribution were acquired by varying the ratio of AgNO$_3$ (aq) relative to NaBH$_4$ (aq) in the final mixture. The samples studied in this work are presented in Table \ref{tab:samples}.

\begin{table}
\begin{tabular}{rcccccccc}
\hline
Sample   & 2 & 4 & 6 & 8 & 10 & 25 & 35 & 45\\ 
\hline
Ratio AgNO$_3$   & 2 & 4 & 6 & 8 & 10 & 25 & 35 & 45\\
      NaBH$_4$   & 25 & 25 & 25 & 25 & 25 & 25 & 25 & 25\\
\hline
\end{tabular}
\caption{An overview of the samples studied in this work. The molar ratio is the varying amount of AgNO$_3$ (aq) relative to NaBH$_4$ (aq) in the final mixture.}\label{tab:samples}
\end{table}

TEM samples were prepared by adding a drop of the final solution on a holy carbon film supported by a Cu grid. The Ag nanocrystals were observed by High-Resolution Transmission Electron Microscopy (HRTEM) and Electron Dispersive Spectroscopy (EDS) in a 200 keV JEOL 2010F microscope with a Gatan imaging filter and detector, and a NORAN Vantage DI+ EDS system. The spherical (Cs) and chromatic aberration (Cc) coefficients of the objective lens were 0.5 mm and 1.1 mm, respectively. The point to point resolution was 0.194 nm at Scherzer focus ($-42$ nm). A probe current of about 0.5 nA at a probe diameter of 1.0 nm can be obtained. 

The size measurements were performed manually on HRTEM images.
In order to relate the material properties of Ag nanocrystals to their use in solar cell applications, the scattering of light from Ag nanocrystals of various sizes was simulated when embedded in SiO$_2$, SiN, Si, and Al$_2$O$_3$ matrices. Assuming spherically shaped nanocrystals, the simulations were carried out using the computer simulation program MiePlot v.4.2.02 \cite{mie:plot}, based on the Mie theory as presented by Bohren and Huffman \cite{bohren:bok}.

\section{Theoretical Discussion of fitting a distribution function}
\label{section:lognormal}
The normal distribution is not appropriate for describing size distributions, as it assigns a positive probability to ``negative sizes''. Several alternative distributions are in use, but the log-normal distribution seems to be a particularly popular alternative.

As early as 1879, Galton and McAlister gave general arguments why sizes are prone to be log-normally distributed 
\cite{log:intro, McAlister:intro}. These arguments are based upon the assumption that sizes (and other physical quantities observable by the senses) are best described by their geometric means, rather than by arithmetic means. From this, Galton and McAlister conclude that one should expect the logarithms of the sizes to be normally distributed.
 
In the first half of the 20th century, several papers were published in which particle size distributions from various contexts were modelled by log-normal distributions \cite{Wightman:intro, Drinker:intro, log:3, Krumbein:intro, Austin:intro, Kottler:intro}, from analyzing the sizes of photographic emulsion grains to those of pulvarized silica.
In line with Galton and McAlister's observations, Kottler inferred that particle size distributions resulting from growth processes can often be assumed to be log-normal, again by arguing that the logarithm of the particle sizes is most likely to be normally distributed \cite{Kottler:intro2}. 

A \emph{log-normal distribution} is defined as a probability distribution of random variables whose logarithm is normally distributed \cite{log:2}. Under general conditions, the central limit theorem implies that the product of independent, identically distributed, positive random variables is asymptotically log-normal. As a consequence, a wide range of frequency distributions can be accurately modelled by a log-normal distribution. For a gentle overview of applications of the log-normal distribution across the sciences, see Limpert, Stahel and Abbt \cite{Limpert:intro}.

The log-normal distribution with parameters $\mu$ and $\sigma$ has probability density function

\begin{equation}
\label{eq:1}
p(x; \mu, \sigma) = \frac{1}{x \sigma \sqrt{2 \pi}} e^{ - \frac{ (\ln x- \mu )^2 }{2 \sigma^2} } ,\qquad x>0.
\end{equation}
The parameters $\mu$ and $\sigma$ coincide with the mean and standard deviation of the natural logarithm of the random variable $X$, which is by definition normally distributed. The random variable $X$ itself has mean
\[ e^{\mu + \sigma^2/2}\]
and standard deviation 
\[ e^{\mu + \sigma^2/2} \sqrt{e^{\sigma^2} - 1}. \]

Given a sample of $N$ nanocrystals with diameters $x_1,\ldots,x_N$, the parameters $\mu$ and $\sigma$ of the best fitting log-normal distribution can be estimated by the
\emph{maximum likelihood estimators}

\[
\widehat{\mu} = \frac{1}{N}\sum_{k=1}^N \ln x_k,\qquad
\widehat{\sigma}^2 = \frac{1}{N}\sum_{k=1}^N (\ln x_k - \widehat{\mu})^2.
\]

Plotting, for each sample, the mean nanocrystal size versus the AgNO$_3$/NaBH$_4$ molar ratio suggests that there exists a linear relation between these two quantities. For a set of uncorrelated data with different uncertainties, as is the case here, it is unreasonable to assume that every data point should be treated equally in estimating an underlying linear relation. Instead, a \emph{best linear unbiased estimator} is obtained by weighing each data point by the reciprocal of its variance \cite{1:1}.

Let us briefly describe this \emph{weighted least square} method in our case. Suppose we have computed, for each sample with ratio $r_i$, the above maximum likelihood estimators $d_i := \widehat{\mu}_i$ of the mean nanocrystal diameter and $\widehat{\sigma}^2_i$ of the variance. The parameters $a$ and $b$ of the purported linear relation $d = ar + b$ are obtained by minimizing the weighted sum of squared residuals

\[ S := \sum_i \frac{1}{\widehat{\sigma}^2_i} (d_i - ar_i - b)^2\]
for $a$ and $b$. In other words, inverting the resulting linear system with equations $\frac{\partial S}{\partial a} = 0$ and $\frac{\partial S}{\partial b} = 0$ provides us with the weighted linear square estimate of the parameters $a$ and $b$.

\section{Results and Discussion}

\subsection{Ag nanocrystal size, distribution, and fitting}

\begin{figure}
  \begin{center}
    \includegraphics[width=0.455\textwidth]{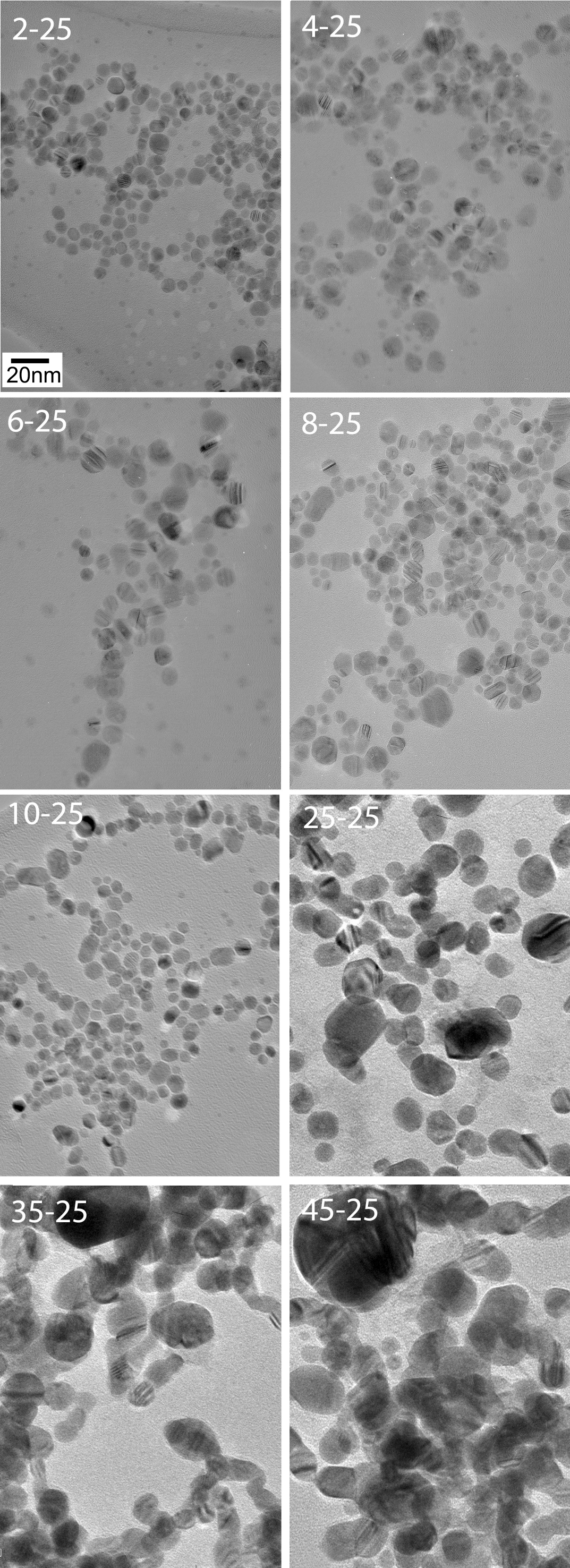}
   \caption{TEM images of Ag nanocrystals in samples 2, 4, 6, 8, 10, 25, 35, and 45.}
    \label{figure:tem}
  \end{center}
\end{figure}

TEM images of Ag nanocrystals dispersed on a carbon film for samples with a ratio of 2, 4, 6, 8, 10, 25, 35, and 45 AgNO$_3$ per 25 units of NaBH$_4$ are shown in Figure \ref{figure:tem}. The Ag nanocrystals in sample 2 seem to be of roughly the same size and spherically shaped. The TEM images show that upon increasing the AgNO$_3$/NaBH$_4$ ratio, the mean Ag nanocrystal size increases, the variation in nanocrystal sizes increases, and the nanocrystals become slightly less spherical. In particular, in the samples with higher ratios, many of the nanocrystals have defects, like twins, as seen from the vertical lines in the nanocrystals. This has been presented in detail in a previous paper \cite{annett:mrs}.   

\begin{figure}
  \begin{center}
    \includegraphics[width=0.5\textwidth]{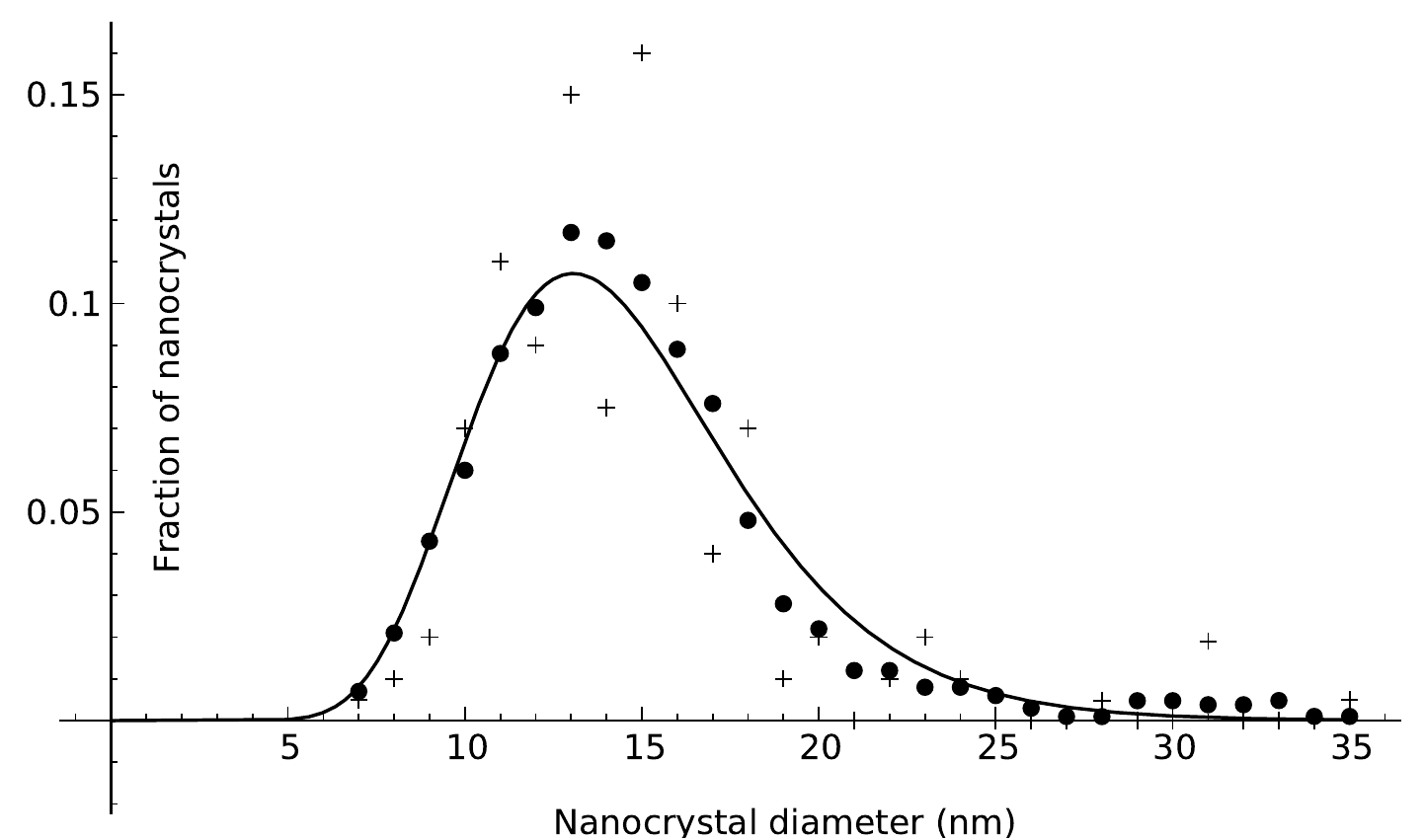}
   \caption{For sample 25, the figure shows a plot of the relative number of nanocrystals of a given size $+$, the associated central moving averages $\bullet$, and a fitted log-normal distribution.}
    \label{figure:SizeDistribution25}
  \end{center}
\end{figure}

The size distributions of the Ag nanocrystals in the different samples were found by manually measuring the nanocrystal diameter on the HRTEM images. The size distribution of sample 25 is presented in Figure \ref{figure:SizeDistribution25}. As this figure shows, the overly erratic behaviour of the raw data points $(i, y_i)$ represented by the $+$ markers does not directly show the trend of the data. A common technique to smoothen such data is to replace, for some natural number $k$, each data point $y_i$ by the \emph{central moving average}

\[\frac{y_{i-k} + \cdots + y_{i-1} + y_i + y_{i+1} + \cdots + y_{i+k}}{2k + 1}.\] 
By replacing each data point by the average of its surrounding points, random fluctuations in the nanocrystal sizes will be canceled. One advantage of this method -- for instance compared to \emph{grouping} -- is that no choice of a starting point needs to be made, avoiding a potential bias in estimating the mean.

The resulting points are represented by the $\bullet$ markers and can be seen to nicely fit a log-normal distribution. For the ratios 2,4,6,8,10,25,35,45 of AgNO$_3$ per 25 units of NaBH$_4$, we have chosen respectively $k = 1,1,1,1,2,2,2,3$.

\begin{figure}
  \begin{center}
    \includegraphics[width=0.5\textwidth]{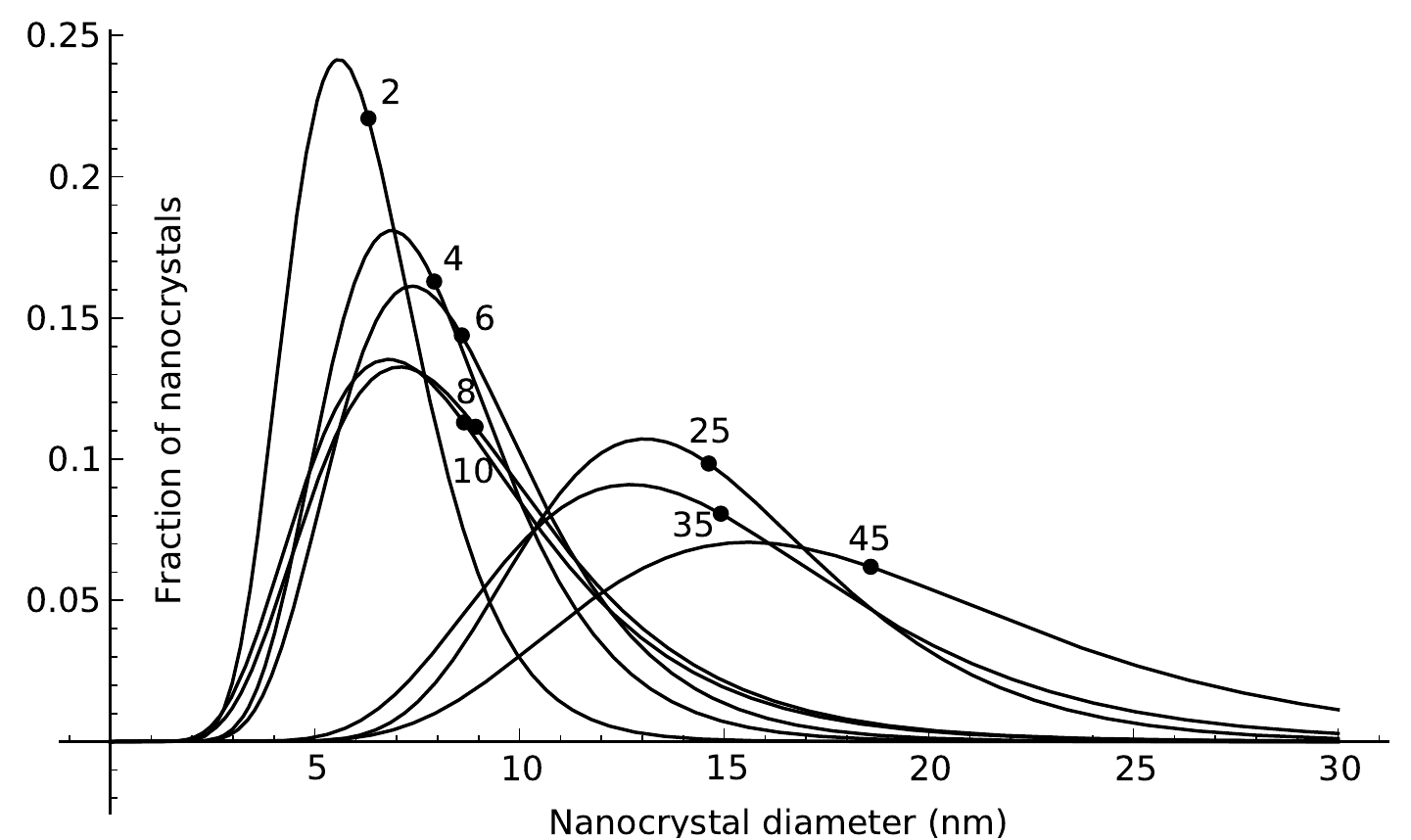}    
   \caption{The fitted log-normal size distributions of Ag nanocrystals from samples 2, 4, 6, 8, 10, 25, 35, and 45. For each fit, the $\bullet$ marks the location of the mean nanocrystal diameter. }
    \label{figure:distribution}
  \end{center}
\end{figure}

In order to compare the size distributions of the different samples, the best fitting log-normal size distribution of the Ag nanocrystals from the samples 2, 4, 6, 8, 10, 25, 35, and 45 are plotted together in Figure \ref{figure:distribution}. In each fitted distribution, the $\bullet$ marks the location of the mean nanocrystal diameter. Numerical values of the estimated parameters $\widehat{\mu}$ and $\widehat{\sigma}$, the mean, and the standard deviation of each distribution can be found in Table \ref{tab:MeanNanocrystalDiameters}. The figure and table show that increasing the ratio AgNO$_3$/NaBH$_4$ results in an increased mean and standard deviation of the associated log-normal nanocrystal size distribution.

\begin{table}
\begin{tabular}{lcccccccc}
\hline
sample             & 2     & 4     & 6     & 8     & 10    & 25    & 35    & 45    \\ \hline
$\widehat{\mu}$    & 1.801 & 2.022 & 2.100 & 2.112 & 2.077 & 2.645 & 2.650 & 2.864 \\
$\widehat{\sigma}$ & 0.284 & 0.306 & 0.319 & 0.393 & 0.400 & 0.274 & 0.327 & 0.342 \\
mean               & 6.307 & 7.915 & 8.590 & 8.925 & 8.644 & 14.62 & 14.92 & 18.58 \\
s.d.               & 1.826 & 2.478 & 2.810 & 3.650 & 3.603 & 4.090 & 5.011 & 6.540 \\
\hline 
\end{tabular}
\caption{For each sample, the table lists the estimated parameters $\widehat{\mu}$ and $\widehat{\sigma}$, the mean, and the standard deviation of the best fitting log-normal distribution.}\label{tab:MeanNanocrystalDiameters}
\end{table}

\begin{figure}
  \begin{center}
    \includegraphics[width=0.5\textwidth]{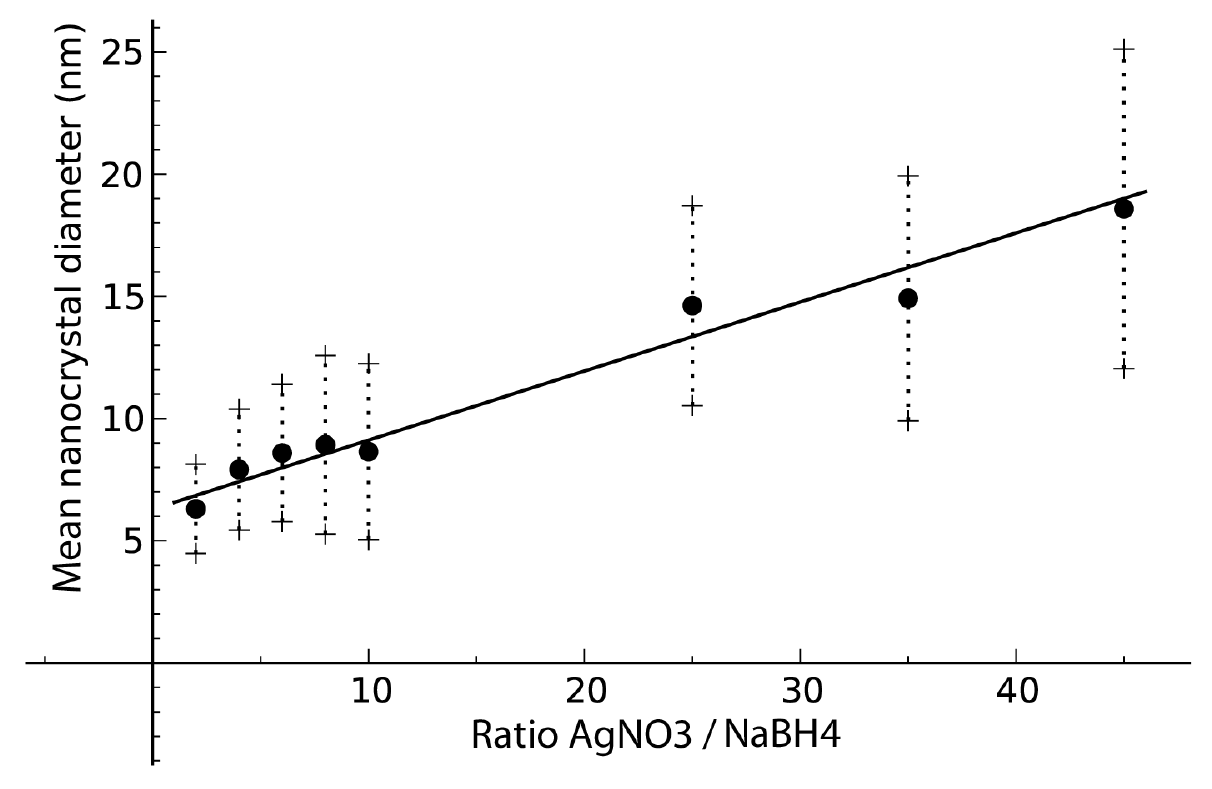}    
   \caption{Mean nanocrystal diameter versus the ratio of AgNO$_3$ per 25 units of NaBH$_4$, with a linear fit.}
    \label{figure:LinearRelationAverages}
  \end{center}
\end{figure}

Figure \ref{figure:LinearRelationAverages} has been made by plotting, for each ratio AgNO$_3$/NaBH$_4$, the mean diameter in the associated log-normal nanocrystal size distribution, the error bars extending one standard deviation in each direction. Using the weighted least squares method described in Section \ref{section:lognormal}, the weights being the reciprocals of the variances of the distributions, the data were fitted by the linear relation $y = 0.283x + 6.290$. From this linear fit, one can estimate the mean nanocrystal size for every ratio and vice versa.

\subsection{Scattering efficiency of the Ag nanocrystals in various matrices}

In order to relate the material properties studied in the previous section to the scattering properties when embedded into a solar cell device, simulation using the program MiePlot v.4.2.02 \cite{mie:plot} was performed. SiN, SiO$_2$, and Al$_2$O$_3$ are the most common materials for passivating the Si surface. The scattering of light from Ag nanocrystals with different sizes embedded in Si, SiO$_2$, SiN, and Al$_2$O$_3$ matrices was simulated and plotted, assuming mono-disperse spherical nanocrystals.

\begin{figure}
  \begin{center}
    \includegraphics[width=0.5\textwidth]{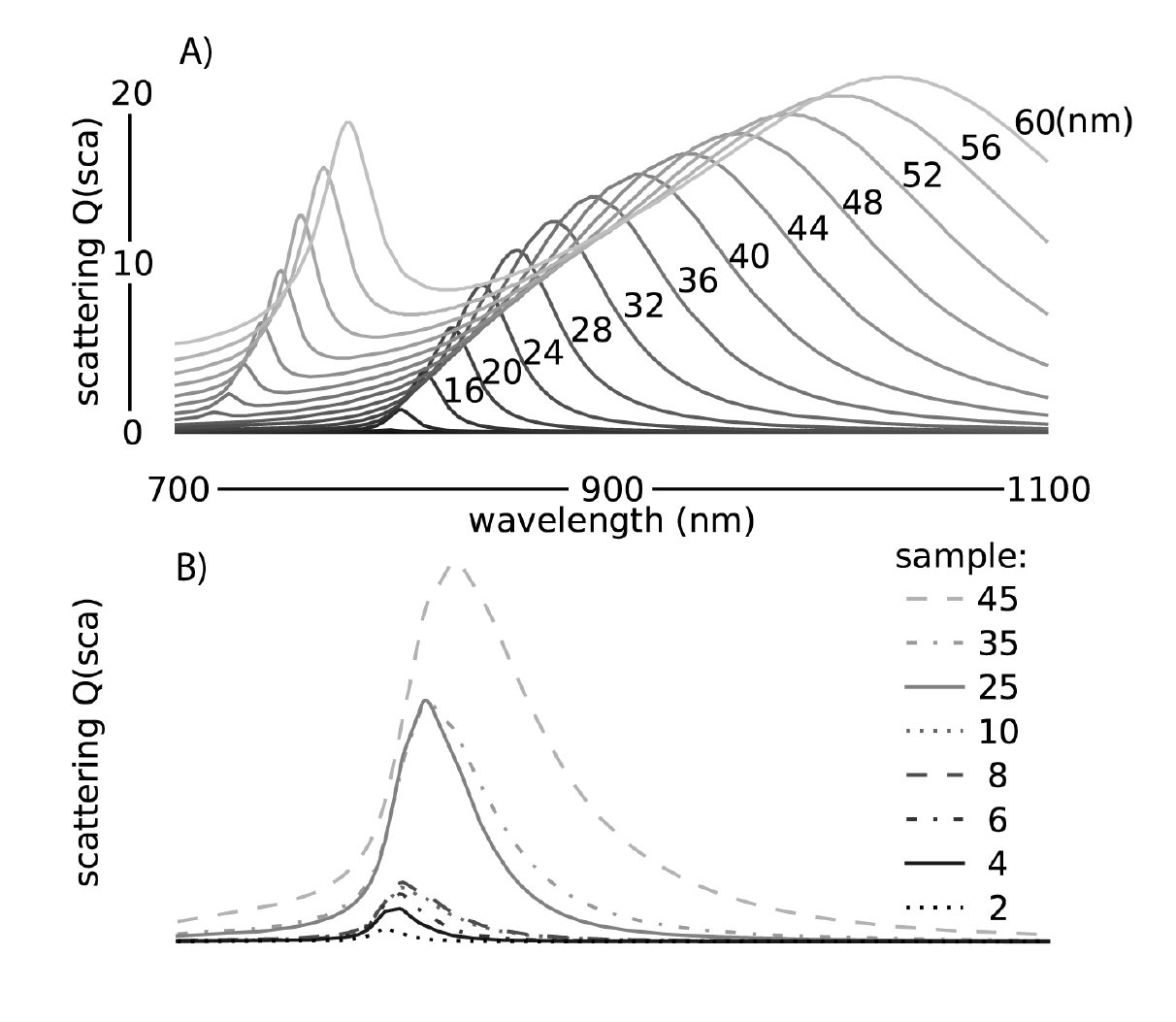}
   \caption{Scattering versus the wavelength of light.
   A) The figure shows these spectra for individual Ag nanocrystal sizes of 4-60 nm in diameter embedded in an Si matrix.
   B) For every fitted size distribution, a corresponding light scattering spectrum was obtained by taking a weighted linear combination of the simulated spectra at various nanocrystal sizes, the weights being the densities of the nanocrystal sizes in the fitted distributions in Figure \ref{figure:distribution}.}
    \label{scattering:Si}
  \end{center}
\end{figure}

Figure \ref{scattering:Si}A shows the scattering versus the wavelength of light for nanocrystals with a nanocrystal size of 4-60 nm in diameter, embedded in an Si matrix. The simulation shows that the location of the peak moves to a larger wavelength as the nanocrystal diameter increases. In addition, the standard deviation increases and a second scattering peak emerges at a smaller wavelength. 

These simulations were performed with the assumption of mono-disperse spherical nanocrystals. As shown in the previous section, the sizes of the nanocrystals in each sample are log-normally distributed. For every fitted size distribution, a corresponding light scattering spectrum was obtained by taking a weighted linear combination of the simulated spectra at various nanocrystal sizes, the weights being the densities of the nanocrystal sizes in the fitted distribution. The resulting spectra are shown in Figure \ref{scattering:Si}B. The main peak of every such spectrum is less symmetric and has a larger standard deviation than corresponding samples with only mono dispersive nanocrystals. Each main peak clearly is a superposition of a multitude of peaks resulting from nanocrystals of other sizes than the mean size.

Sample 45 has a mean nanocrystal diameter of 18.6 nm. Comparing the simulated scattering spectrum of a mono-disperse spherical nanocrystal with a diameter of 20 nm to the weighted spectrum of sample 45, one finds that the peaks are located at roughly the same wavelength. However, the width and shape of the peaks are different, and one can distinguish three additional peaks around the main peak of the weighted spectrum. This shows that the size distribution of the Ag nanocrystals when embedded in a solar cell has a large effect upon the scattering properties of the device.

\begin{figure}
  \begin{center}
    \includegraphics[width=0.5\textwidth]{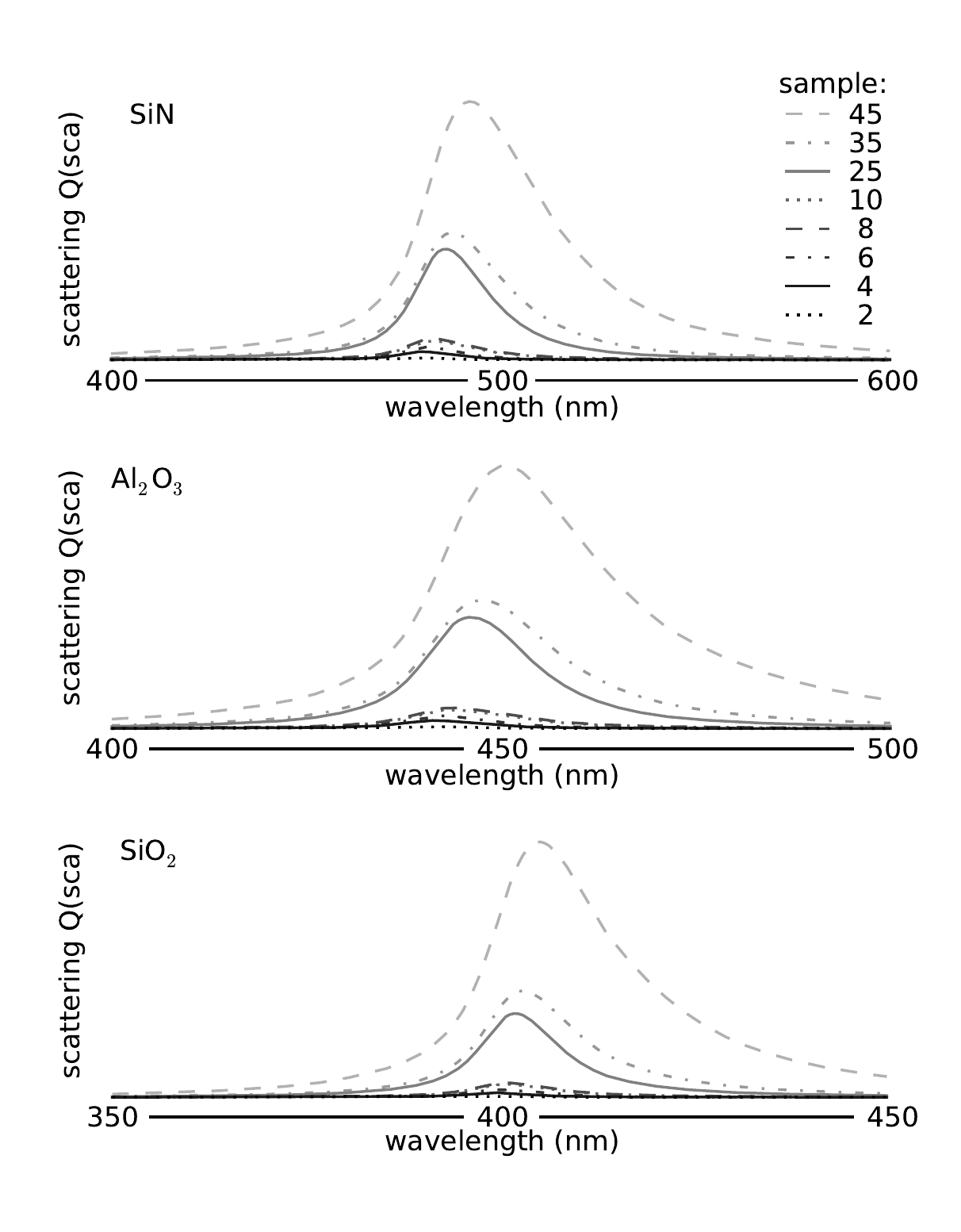}
   \caption{Light scattering spectra of Ag nanocrystals in an SiO$_2$, SiN, and Al$_2$O$_3$ matrix, obtained as described in the caption of Figure \ref{scattering:Si}B.}
    \label{scattering:scatteringlog}
  \end{center}
\end{figure}

A similar analysis was carried out for Ag nanocrystals in SiO$_2$, SiN, and Al$_2$O$_3$ matrices (see Figure \ref{scattering:scatteringlog}). The same effects as described for Ag nanocrystals in an Si matrix are shown for the Ag nanocrystals embedded in these matrices. However, for these matrices the maximum scattering peaks lie in the range of 400-500 nm. 

\begin{figure}
  \begin{center}
    \includegraphics[width=0.4\textwidth]{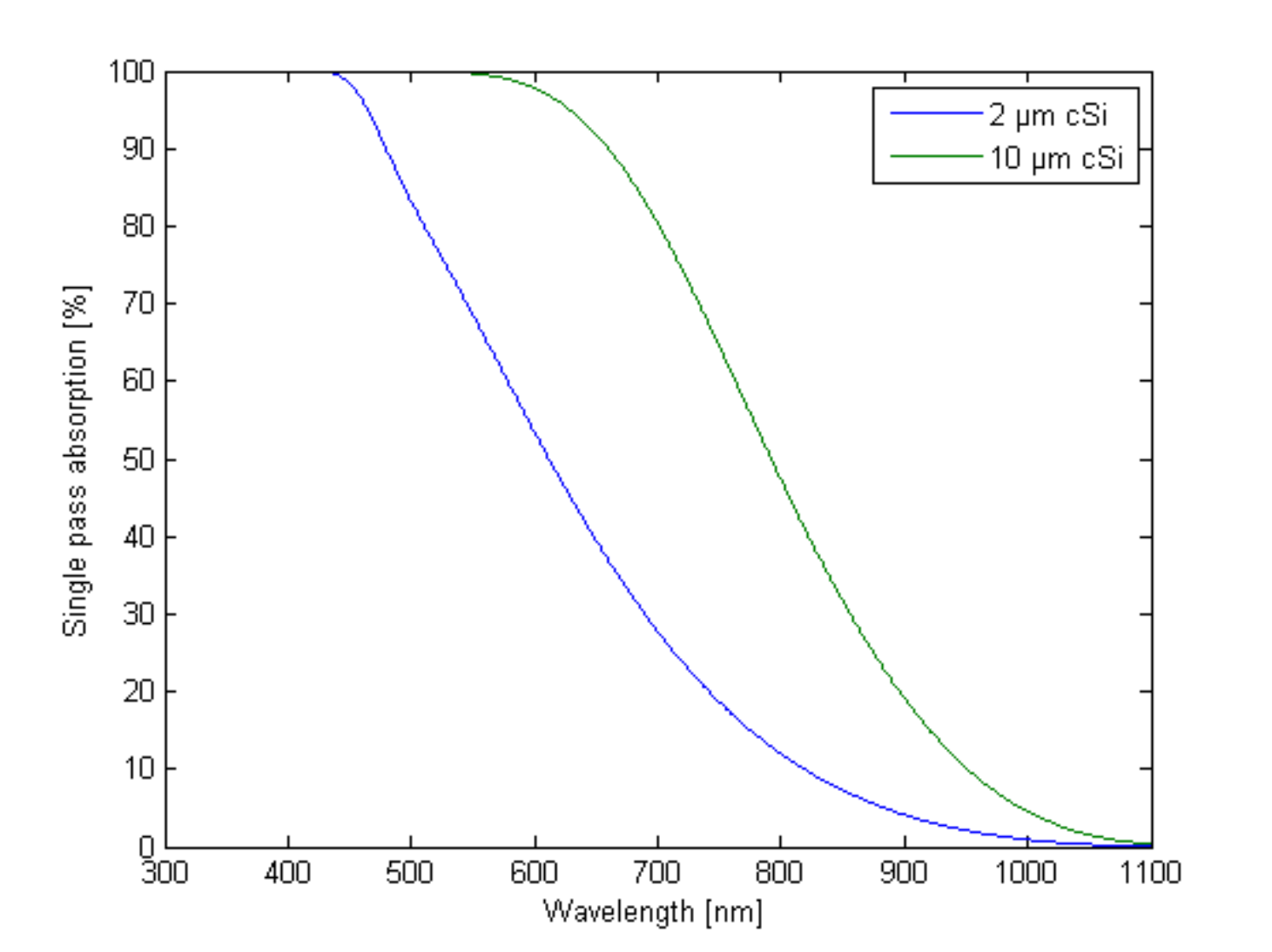}    
   \caption{Single pass absorption in a 2 $\mu$m and 10 $\mu$m Si solar cell. The plot shows how much of the light is absorbed before the light reaches the back side.}
    \label{scattering:jo}
  \end{center}
\end{figure}

The solar spectrum convertible by crystalline silicon cells lies between 370 and 1100 nm (100 nm corresponds to $\sim$1.1 eV, which is the band gap of Si) \cite{Kuznicki:solar}. It is desirable to have a broad scattering peak at a wavelength where the Si absorbs most of the light. When the thickness of the Si wafer decreases, more of the large wavelength light (600-1100 nm) will pass through the solar cell and will not be absorbed. Figure \ref{scattering:jo} shows single pass absorption in a 2 $\mu$m and 10 $\mu$m Si solar cell. The figure shows how much of the light is absorbed before the light reaches the back side. When the thickness of the Si wafer decreases, less of the large wavelength light will be absorbed. Red-shifting the scattering peak to these wavelengths may therefore result in more absorption in the solar cell, resulting in an increased efficiency.  

\begin{figure}
  \begin{center}
    \includegraphics[width=0.4\textwidth]{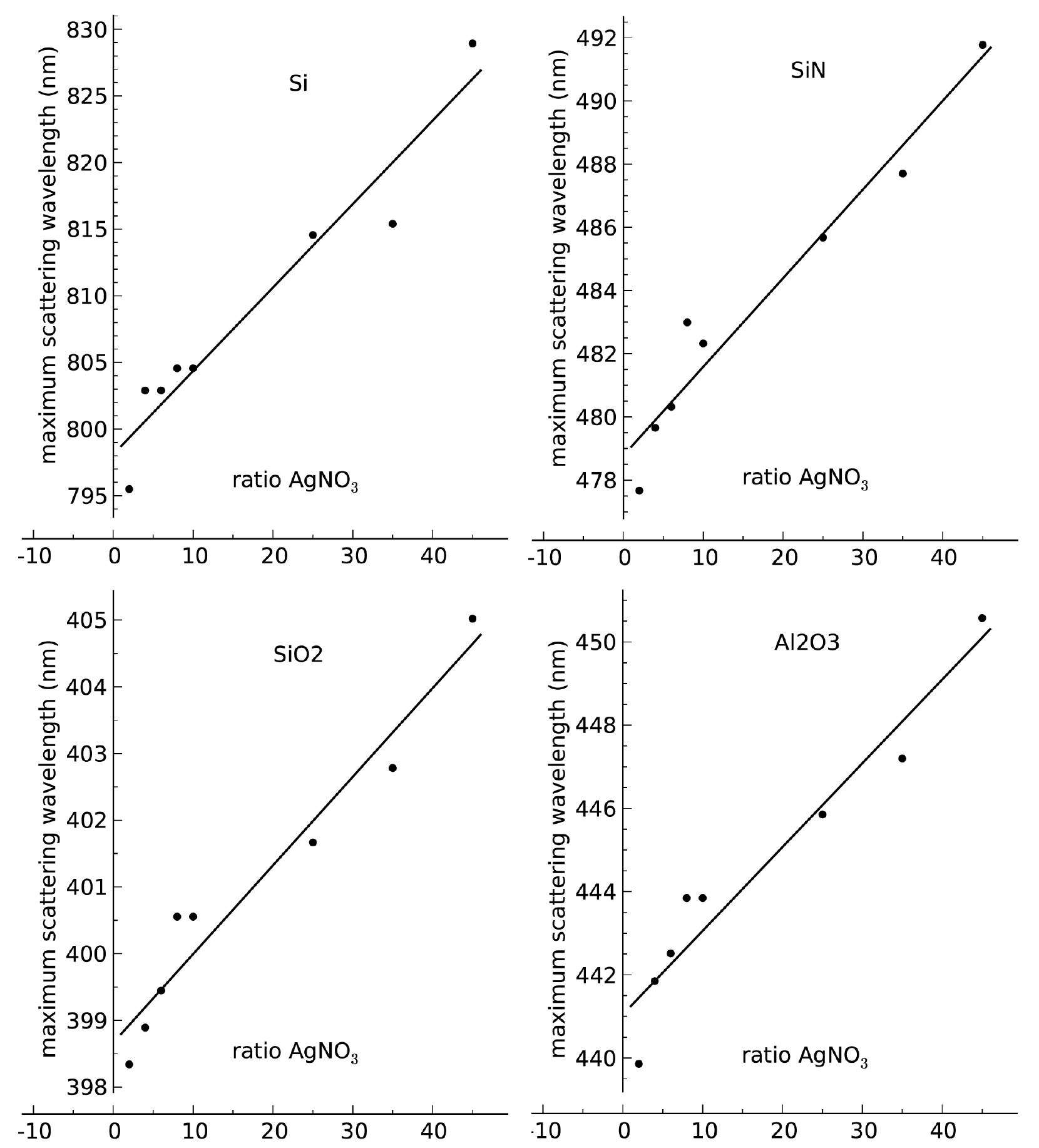}
   \caption{The wavelength of maximum scattering versus the molar ratio for Ag nanocrystals in a Si, SiO$_2$, SiN, and Al$_2$O$_3$ matrix.}
    \label{scattering:rvms}
  \end{center}
\end{figure}

In order to relate the wavelength with maximum scattering to the corresponding AgNO$_3$/NaBH$_4$ ratio, these quantities were plotted in Figure \ref{scattering:rvms}. The figure shows the wavelength with maximum scattering versus the molar ratio for Ag nanocrystals embedded in Si, SiO$_2$, SiN, and Al$_2$O$_3$ matrices. These plots can be used to find the molar ratio needed to make nanocrystals that scatter a particular wavelength maximally.

From the solar energy spectrum presented by Atwater and Polman \cite{Atwater:solar} it seems useful to have a scattering maximum located at a wavelength of at least 600 nm, for example at 700 nm. A maximum scattering peak at 700 nm seems possible only if the Ag nanocrystal size is 1060 nm, 608 nm, and 352 nm when embedded in a SiO$_2$, Al$_2$O$_3$, and SiN matrix, respectively. These sizes were found by extrapolating the wavelength of maximum scattering versus the molar ratio for Ag nanocrystals in Figure \ref{scattering:rvms} for the three materials. The surface plasmon resonance depends on the nanocrystal size being smaller than the wavelength of incoming light. Since most of the absorbed light has a wavelength of 400 nm or larger, only nanocrystals with diameter at most 400 nm will yield a surface plasmon resonance effect. This makes SiN the best material for embedding the Ag nanocrystals.

It follows that the Ag nanocrystals studied in this article are much smaller than what is most suitable for maximum scattering of light with a wavelength of 700 nm. Nevertheless, some additional absorption of low wavelength light is possible to obtain when depositing these Ag nanocrystals on a solar cell device.

\section{Conclusion}

Silver nanocrystals were made by chemical reduction synthesis of silver salts by sodium borohydride. The nanocrystal size distributions were fitted with log-normal distributions. The plotted distributions showed that increasing the AgNO$_3$/NaBH$_4$ molar ratio resulted in an increased mean nanocrystal size, and a linear relation between the mean nanocrystal size and the AgNO$_3$/NaBH$_4$ molar ratio was found. By simulating scattering spectra using Mie plot and weighing these by the log-normal size distribution densities, scattering spectra were found for each sample. For a sample of Ag nanocrystals whose sizes are log-normally distributed, the scattering peak is less symmetric and broader than for nanocrystals of a fixed size. In order to increase the efficiency of a solar cell device using a thin Si wafer, Ag nanocrystals embedded in a SiN layer with a diameter of 352 nm should be made.

\section{Acknowledgements}

We wish to thank Jeyanthinath Mayandi for guiding Jack Bonsak through the process of making the Ag nanocrystals, Paul Kettler for suggesting to use a log-normal distribution to describe the nanocrystal sizes, Jo Gjessing for making the absorption spectra, and Erik Marstein for his insightful comments on a draft of this paper. The project has been funded by The Norwegian Research Council project number 185414.


\end{document}